\documentclass[preprint,aps,amsmath,showpacs,pre]{revtex4}
\usepackage{graphicx}
\usepackage{amssymb}
\usepackage{amsmath}
\usepackage{dcolumn}
\usepackage{comment}
\usepackage{color}
\DeclareMathOperator{\arccosh}{arch}

\begin{document}
\title{Entropy of fully-packed rigid rods on generalized Husimi trees: a route to the square lattice limit}
\author{Nathann T. Rodrigues}
\email{nathan.rodrigues@ufv.br}
\affiliation{Instituto de F\'isica and National Institute of Science and Technology for Complex Systems, Universidade Federal Fluminense, Avenida Litor\^anea s/n, 24210-346 Niter\'oi, Rio de Janeiro, Brazil} \author{Tiago J. Oliveira}
\email{tiago@ufv.br}
\affiliation{Departamento de F\'isica, Universidade Federal de Vi\c cosa, 36570-900, Vi\c cosa, Minas Gerais, Brazil}
\author{J\"urgen F. Stilck}
\email{jstilck@id.uff.br}
\affiliation{Instituto de F\'isica and National Institute of Science and Technology for Complex Systems, Universidade Federal Fluminense, Avenida Litor\^anea s/n, 24210-346 Niter\'oi, Rio de Janeiro, Brazil} 

\date{\today}

\begin{abstract}
Although hard rigid rods ($k$-mers) defined on the square lattice have been widely studied in the literature, their entropy per site, $s(k)$, in the full-packing limit is only known exactly for dimers ($k=2$) and numerically for trimers ($k=3$). Here, we investigate this entropy for rods with $k \le 7$, by defining and solving them on Husimi lattices built with diagonal and regular square lattice clusters of effective lateral size  $L$, where $L$ defines the level of approximation to the square lattice. Due to an $L$-parity effect, by increasing $L$ we obtain two systematic sequences of values for the entropies $s_L(k)$ for each type of cluster, whose extrapolations to $L \rightarrow \infty$ provide estimates of these entropies for the square lattice. For dimers, our estimates for $s(2)$ differ from the exact result by only $0.03\%$, while that for $s(3)$ differs from best available estimates by $3\%$. In this paper, we also obtain a new estimate for $s(4)$. For larger $k$, we find that the extrapolated results from the Husimi tree calculations do not lie between the lower and upper bounds established in the literature for $s(k)$. In fact, we observe that, to obtain reliable estimates for these entropies, we should deal with levels $L$ that increase with $k$. However, it is very challenging computationally to advance to solve the problem for large values of $L$ and for large rods. In addition, the exact calculations on the generalized Husimi trees provide strong evidence for the fully packed phase to be disordered for $k\geq 4$, in contrast to the results for the Bethe lattice wherein it is nematic, thus providing evidence for a high density nematic-disordered transition in the system of $k$-mers with vacancies.
\end{abstract}


\maketitle

\section{Introduction}
\label{SecI}

The problem of a phase transition to an ordered phase in a system of long cylindrical
rods in solution, with excluded volume interactions only, was considered by
Onsager, who showed that a solution of long rods would undergo a transition
between an isotropic and a nematic ordered state as the increasing density passes through a critical value \cite{o49}. Such ordered phases were also found in approximate
calculations of systems of semi-flexible polymers in solution \cite{f56} if
the chains are sufficiently stiff. The case of rods with rectangular cross-sections and discrete orientations 
in the three-dimensional continuous space was studied also \cite{z63}. 
A review of these models and their properties may be found in \cite{vl92}. 
In two dimensions, with continuous orientations and positions, it is known that 
the system does not order, but undergoes a Kosterlitz-Thouless transition between a low-density phase with exponential decay of correlations to a high-density one where the correlations decay with a power law \cite{s71,f85,k05,v09}.

In the related lattice model, rods are formed by $k$ consecutive sites along one of the directions of the edges, called $k$-mers. The particular case of dimers ($k=2$) has a long history, it can be shown that the orientational  correlations of the rods decay exponentially with the distance between them if the dimers do not occupy all sites of the lattice, and in the full lattice limit they decay with a power law tail for all dimensions $d \ge 2$ \cite{hl72,gdj07}. In a seminal paper, for general $k$, Ghosh and Dhar \cite{gd07} studied the model with vacancies on the square lattice, using grand-canonical simulations and theoretical arguments in the large $k$ limit. They found out that, while the system is always in an isotropic phase for $k\le 6$, a continuous transition to a nematic phase, where the rods are preferentially in one of the two directions, happens at sufficiently high density of rods with $k \ge 7$. Moreover, at even higher densities, close to the full-packing limit, where simulations are difficult due to jamming, it was argued that a reentrant transition to an isotropic phase must be present. Additional simulations provided evidence that the first transition is in the Ising universality class for rods on the square lattice and in the three-state Potts universality class when they are placed on the triangular lattice \cite{rp08}. Even using a new Monte Carlo scheme which reduces the long relaxation times in the high density region and was introduced in Ref. \cite{k13}, which leads to more precise results for the second transition, its universality class is still not clear, and actually recent results show that it is actually discontinuous for large $k$ \cite{sdr21}.

In the full lattice limit the entropy of dimers ($k=2$) on the square lattice was calculated exactly a long time ago, using pfaffians \cite{k61} and transfer matrices \cite{l67}, the result for the entropy per site is $s(2)=S(2)/N=G/\pi = 0.29156...$, where $G$ is Catalan's constant and $N$ is the number of sites in the lattice. A recent summary of the generalisations of Lieb's transfer matrix calculations \cite{l67} may be found in \cite{np21}.
Precise transfer matrix estimates of the entropy in this limit for trimers ($k=3$) were obtained in \cite{gdj07}, leading to $s(3)=0.158520 \pm 0.000015$. 
However, for larger values of $k$, besides very recent estimates provided by simulations \cite{p21}, accurate estimates of $s(k)$ are still missing, to the best of our knowledge. There exist, however, interesting results for the lower and upper bounds of $s(k)$ in the literature in the full square lattice limit.  For instance,  the lower bound $s(k) \ge \frac{4G}{\pi k^2}$ was established in Ref. \cite{gp79}, where the upper limits $s(k) \le \frac{1}{k^2} \ln\left(\frac{k}{2}\right) + \frac{4G}{\pi k^2}$ for even $k$ and $s(k) \le \frac{1}{k^2} \ln\left(\frac{k-1}{2}\right) + \frac{1}{\pi k^2} \int_0^{\pi} \arccosh \left(\frac{2k}{k-1} - \cos\phi \right)d\phi$ for odd $k$ were also obtained, with $G$ being Catalan's constant. Very recently, the asymptotic behavior of this entropy for large $k$ was studied in Ref. \cite{dr21}. Besides providing better lower bonds as compared to the one above, it was shown that it approaches $s(k) = k^{-2} \ln k$ for $k \gg 1$. This result was extended to hypercubic lattices also.

The behavior of the model on the Bethe lattice (the core of an infinite Cayley tree), with arbitrary even coordination number $q$, coverage and rod length $k$, was studied in \cite{drs11}. In order to check for possible surface effects inherent to the Cayley tree, which could be particularly relevant at full-packing, the rods were analyzed in \cite{drs11} also on a random graph where all sites have the same coordination number \cite{b94}, such that the surface is absent. The exact solutions of models on this \textit{random locally treelike layered} (RLTL) lattice usually correspond to the ones which follow from the Bethe approximation and, in fact, equivalent results were found for the rods on the Bethe and RLTL lattices \cite{drs11}. If the infinite excluded volume repulsion is relaxed, replaced by statistical weights for multiply occupied sites, it was found that both transitions [isotropic-nematic-isotropic] may appear for the model on the RLTL lattice \cite{kr13}. Of particular interest here are the results for the hard-core problem in coordination $q=4$, where a continuous isotropic-nematic transition was found in these lattices already for $k \ge 4$ \cite{drs11}. We remark that the ordering of rods smaller than the smallest ones which lead to a nematic phase on the square lattice is indeed expected, since the Bethe lattice solution is equivalent to a mean-field approximation \cite{b82} and, because of this, the critical exponents are classical and ordered phases may appear in situations where they are absent in better approximations or exact results. The second transition, from the nematic to the high density isotropic phase is absent on the Bethe lattice solution. Moreover, the entropies at full-packing are considerably smaller than those for the regular lattice. For example, $s(2) = 0.26162$ and $s(3) = 0.05663$ \cite{drs11}, which deviate by 10\% and 64\% from the values above for the square lattice.
These differences lead us to inquire whether solutions on improved treelike lattices may provide more reliable approximations to the behavior of rods on the square lattice. For instance, in a recent paper \cite{Nathann21}, hard square lattice gases were investigated on a sequence of generalized Husimi lattices \cite{h50} (built with diagonal square lattice cells which share $L$ sites with each of their four neighboring cells) and accurate estimates for the critical density and fugacity for the models on the square lattice were obtained from extrapolations to $L \rightarrow \infty$ of the numerically exact results for increasing $L$. This is in agreement with previous findings by Monroe \cite{Monroe}, who introduced and successfully applied this approach to determine the critical parameters of the Ising and other spin models. In this paper, we study fully-packed rods on these Husimi lattices, as well as on another sequence proposed by Kobayashi and Suzuki \cite{ks93}, where the diagonal square cells are replaced by regular square lattice cells that share $L$ sites and $L-1$ edges with neighboring ones. We are interested mainly in two quantities: the entropy per site and the nematic order parameter.  After reviewing the results for the Bethe lattice in this limit \cite{drs11}, we proceed solving the model on the ordinary Husimi lattice, built with elementary squares, which is the core of a square Husimi tree \cite{h50}. We then consider the generalized trees, to obtain two systematic sequences of values of the entropy per site and of the nematic order parameter for $k \le 7$. By extrapolating these values for $L \rightarrow \infty$, accurate estimates of these parameters on the square lattice are obtained for the smaller $k$'s.

The rest of this paper is organized as follows. We start reviewing the results for a Bethe lattice with coordination number $q=4$ in Sec. \ref{SecBL}, proceed to the Husimi lattice built with elementary squares in Sec. \ref{SecHL} and then to sequences of trees with larger cells, with diagonally (Sec. \ref{secDiag}) and regularly (Sec. \ref{secKS}) oriented square clusters. Final discussions and conclusions follow in Sec. \ref{secConc}.

\section{Bethe lattice}
\label{SecBL}

For the sake of completeness, before starting the study of hard rigid rods on Husimi lattices, we summarize the main results for them on the Bethe and RLTL lattices, as obtained in \cite{drs11}, for the full lattice limit. These lattices may be viewed as a tree where the cells are sites. The entropy per site (and in units of $k_B$) of the model in this limit, for lattices with coordination number $q=2d$ and rods with $k$ monomers each is given by \cite{drs11}:
\begin{equation}
s=\sum_{i=1}^d\left[\left(1-\frac{k-1}{k}\rho_i\right)\ln\left(1-\frac{k-1}{k}\rho_i\right)-\frac{\rho_i}{k}\ln\frac{\rho_i}{k}\right],
\label{s_rho}
\end{equation}
where $\rho_i$ is the density of rods in direction $i=1,2,\ldots,d$, such that $\sum_{i=1}^d \rho_i=1$. The actual entropy of the system may then be found by maximizing the entropy \ref{s_rho} over the densities satisfying the constraint of full occupancy. For $k\geqslant 4$, the maximum entropy phase is found to be the one that corresponds to a nematic phase, that is, $\rho_2=\rho_3 = \cdots=\rho_d=\rho_1-\psi$, where $\psi>0$ is the nematic order parameter. This is different from what is observed on the square lattice, where the system is isotropic in the full lattice limit \cite{gd07}. We may then write the entropy as a function of the order parameter:
\begin{eqnarray}
s(\psi)&=&\left\{ 1-\frac{(k-1)[1+(d-1)\psi]}{kd}\right\}\ln\left\{ 1-\frac{(k-1)[1+(d-1)\psi]}{kd}\right\}+\nonumber \\
&&(d-1)\left[ 1-\frac{(k-1)(1-\psi)}{kd}\right]\ln\left[ 1-\frac{(k-1)(1-\psi)}{kd}\right]-\nonumber \\
&&\frac{1+(d-1)\psi}{kd}\ln \left[\frac{1+(d-1)\psi}{kd}\right]-\frac{(d-1)(1-\psi)}{kd}\ln  \left[\frac{1-\psi}{kd}\right]
\label{s_psi}
\end{eqnarray}
and search for its extrema, to find the order parameter in the full lattice limit. Let us recall that for $k=2$ and $3$ the maximum of this entropy is located at $\psi=0$, so that the phase is isotropic. For larger rods ($k \geqslant 4$) a phase transition between an isotropic and a nematic phase happens at lower densities of rods and the system is ordered even at the full lattice limit. In this case, the entropy has a minimum at $\psi=0$, assuming negative values there (see Tabs. \ref{TabEntDiag} and \ref{TabEntKS}, where the Bethe lattice results correspond to $L=0$). The maximum is located at a value of $\psi$ which is given by the equation:
\begin{equation}
[kd-(k-1)(1-\psi)]^{k-1}(1-\psi)-\{kd-(k-1)[1+(d-1)\psi]\}^{k-1}[1+(d-1)\psi]=0.
\end{equation}
The values of these order parameters and the entropies, for $q=4$, are shown in Tabs. \ref{TabEntDiag} and \ref{TabEntKS} for $k \le 7$. As already noticed in the Introduction, the values of $s(2)$ and $s(3)$ are smaller than the square lattice values [$s(2)=G/\pi = 0.29156\ldots$ and $s(3)=0.158520 \pm 0.000015$] by 10\% and $64$\%, respectively.

The asymptotic behavior of $\psi$ at the maximum follows \cite{drs11}:
\begin{equation}
1-\psi \approx \frac{d}{k^{k-1}},\,\,\,\,k \to\infty.
\end{equation}
By substituting for $\psi$ in Eq. \ref{s_psi}, we find the asymptotic entropy (for $k \rightarrow \infty$) on the Bethe lattice as $s(k) = (d-1)k^{-k}$, which, for $d=2$, vanishes much faster than the square lattice result $s(k) = k^{-2}\ln(k)$ \cite{dr21}.

\section{Ordinary Husimi lattice}
\label{SecHL}

\begin{figure}
\centering
\includegraphics[width=7.0cm]{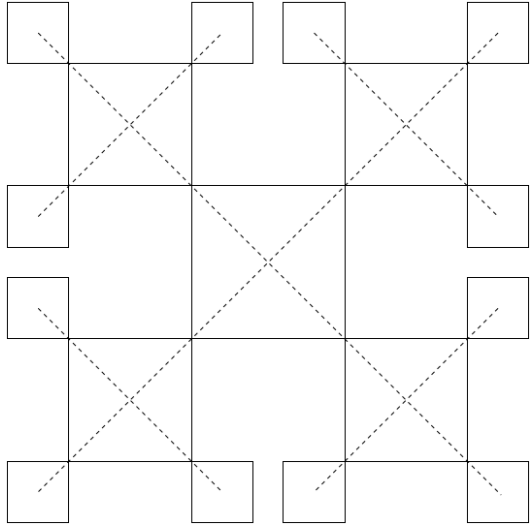}
\caption{A Husimi tree built with squares, with the central square and two additional generations of squares. The dual lattice, represented by dashed lines, is a Cayley tree.}
\label{Fight}
\end{figure}

Let us start analyzing the simplest case of a tree built with clusters, consisting of elementary squares, which is shown in Fig. \ref{Fight}. The solution of a given model in the core of the infinite tree (i.e., in the thermodynamic limit), known as Husimi lattice (HL) \cite{h50}, can be seen as the first level of approximation for its behavior on the square lattice \cite{Monroe,Nathann21}. We may index the levels of approximation by the number of sites shared between each pair of adjacent clusters, which is $L=1$ for this lattice. In the same token, the solution of the model on the Bethe lattice could be seen as a kind of zeroth-level ($L=0$) approximation. 

\begin{figure}
\centering
\includegraphics[width=14.0cm]{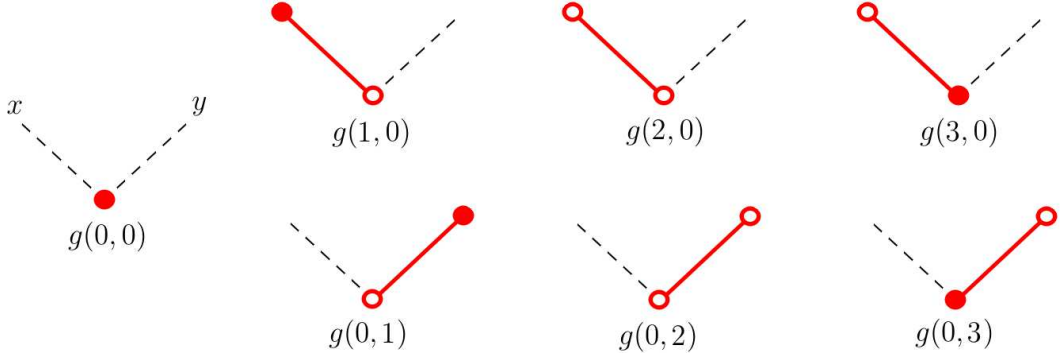}
\caption{Possible configurations for the root site of a HL built with squares for a tetramer ($k=4$). The solid and open circles represent endpoint and internal monomers, respectively, while the solid (red) lines are the bonds connecting them. The directions $x$ and $y$ of the edges are also indicated.}
\label{Fig2}
\end{figure}

\begin{figure}
\centering
\includegraphics[width=14.0cm]{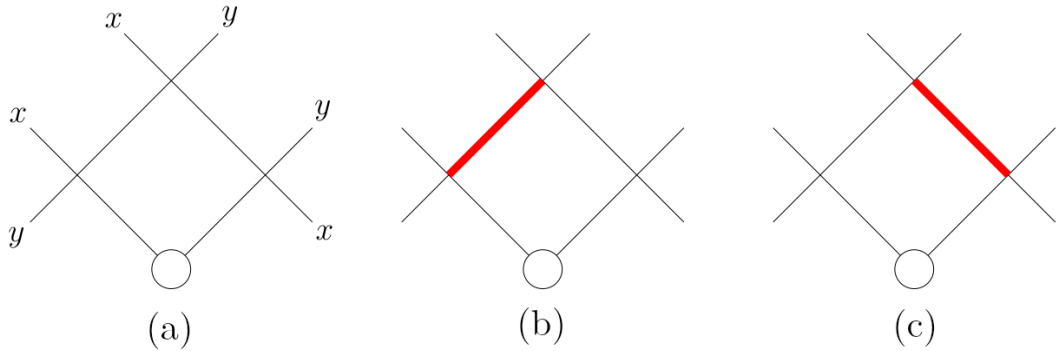}
\caption{Possible configurations of the root square for the root site configuration $(0,0)$. The thick (red) lines indicate edges occupied by $k$-mer bonds. The directions $x$ and $y$ of the edges are also indicated in (a).}
\label{rr00}
\end{figure}

As usual, to obtain the number of configurations of $k$-mers placed on the HL, we start considering rooted subtrees for fixed configurations at the root site. We may label the directions of the edges of the tree by $x$ and $y$ [see Figs. \ref{Fig2} and \ref{rr00}], so that the configuration of the root site may be defined by describing the rods coming from above. As illustrated in Fig. \ref{Fig2} for $k=4$, we may have the state $(0,0)$ if no rods reach the root site; $(i,0)$, for $i=1,2,\ldots,k-1$, if a rod in the $x$ direction with $i$ monomers incorporated reaches it; and $(0,i)$ if the incident rod is in the $y$ direction. Therefore, in general, we have $2k-1$ root configurations, so this is the number of partial partition functions (ppf's) of the subtrees.

A recursion relation for the ppf associated with the configuration $(i,j)$, let us denote it by $g(i,j)$, can be obtained by considering the operation of attaching three subtrees (with $M$ generations each) to a new central square with the root site in configuration $(i,j)$, in order to build a subtree with an additional generation ($M+1$). We will discuss in some detail the derivation of the contributions to $g(0,0)$. It is convenient to order them according to the possible occupations (by $k$-mer bonds) of the two edges of the rooted square which are not connected to root site. For the root configuration $(0,0)$, they can be empty, or occupied by the $i$th bond of a $k$-mer in the $x$ or $y$ direction, as indicated in Fig. \ref{rr00}. Thereby, the contributions to the ppf $g'(0,0)$, in generation $M+1$, are $[g(k-1,0)+g(0,k-1)]^3$, $[g(k-1,0)+g(0,k-1)]\sum_{n=0}^{k-2}g(n,0)g(k-n-2,0)$, and $[g(k-1,0)+g(0,k-1)]\sum_{n=0}^{k-2}g(0,n)g(0,k-n-2)$ for the edge configurations (a), (b), and (c) in Fig. \ref{rr00}, respectively, where the unprimed $g(i,j)$ are in generation $M$. In a similar way, the other recursion relations may then be derived, the result is:
\begin{subequations}
\begin{eqnarray}
  g'(0,0)&=&[g(k-1,0)+g(0,k-1)]^3+[g(k-1,0)+g(0,k-1)]\left[\sum_{n=0}^{k-2}g(n,0)g(k-n-2,0)\right]+\nonumber \\
  &&[g(k-1,0)+g(0,k-1)]\left[\sum_{n=0}^{k-2}g(0,n)g(0,k-n-2)\right],\\
  g'(i,0)&=&g(i-1,0)\{[g(k-1,0)+g(0,k-1)]^2+\nonumber \\
  &&\sum_{n=0}^{k-2}g(n,0)g(k-n-2,0)\},\\
  g'(0,i)&=&g(0,i-1)\{[g(k-1,0)+g(0,k-1)]^2+\nonumber\\
  &&\sum_{n=0}^{k-2}g(0,n)g(0,k-n-2)\},
\end{eqnarray}
\label{eqRRsL1}
\end{subequations}
where $i=1,\ldots,k-1$ and $k \geqslant 2$.
 
These recursion relations (RRs) diverge in the thermodynamic limit (i. e., when $M \rightarrow \infty$), so that it is convenient to work with ratios of them, which are defined as
\begin{equation}
R(i,0) \equiv \frac{g(i,0)}{g(0,0)} \quad \text{and} \quad R(0,j) \equiv \frac{g(0,j)}{g(0,0)},
\end{equation}
for $i,j=1,\ldots,k-1$. Note that the RRs above can be written as $g'(i,j) = [g(0,0)]^3 f(i,j)$, where the functions $f(i,j)$ depend only on the ratios. Thereby, one readily finds RRs for the ratios as $R'(i,0) = f(i,0)/f(0,0)$ and $R'(0,j) = f(0,j)/f(0,0)$. These RRs are convergent and their real, positive and stable fixed points define the stable thermodynamic phases of the $k$-mers on the Husimi tree.

Although the fixed point may be reached iterating the recursion relations \ref{eqRRsL1}, in the present case, due to their rather simple expressions, we can propose the following Ansatz for the fixed point values of the $2k-2$ ratios of the ppf's:
\begin{subequations}
\begin{eqnarray}
R^*(i,0)&=&x_1^i,\\
R^*(0,i)&=&x_2^i.
\end{eqnarray}
\end{subequations}
At the fixed point, the variables $x_1$ and $x_2$ obey the following pair of non-linear equations:
\begin{subequations}
\begin{eqnarray}
x_1[H_{k-1}^3+(k-1)H_{k-1}H_{k-2}]&=&H_{k-1}^2+(k-1)x_1^{k-2},\\
x_2[H_{k-1}^3+(k-1)H_{k-1}H_{k-2}]&=&H_{k-1}^2+(k-1)x_2^{k-2};
\end{eqnarray}
\label{fpe}
\end{subequations}
where $H_{n}=x_1^n+x_2^n$.
For the isotropic fixed point $x_1=x_2=x$, the variable $x$ may be found easily, being given by: 
\begin{equation}
    x=\left[\frac{\sqrt{k^2-2k+2}-k+2}{4}\right]^{1/k}.
\end{equation}
Therefore, the fixed point equations \ref{fpe} have at least this solution, associated with the isotropic phase, for any value of $k$. For large rods, an additional nematic fixed point is also a solution. It is therefore interesting to study the stability of the fixed points, since in general we expect them to be stable to be physically meaningful. If more than one fixed point is stable in some region of the parameter space, the one with the lowest free energy will correspond to the thermodynamically stable phase.  We thus consider the $(2k-2) \times (2k-2)$ jacobian of the recursion relations
\begin{equation}
    J_{\vec{v},\vec{w}}=\frac{\partial R_{\vec{v}}}{\partial R_{\vec{w}}},
\end{equation}
where the derivatives are evaluated at the fixed point and the vectors $\vec{v}$ and $\vec{w}$ denote all the allowed pairs $(i,j)$ in $R(i,j)$. The jacobian matrix is non-symmetric and in general the dominant eigenvalues, $\lambda$, are complex. The modulus of the dominant eigenvalue determines the stability of the fixed point, it will be stable if $|\lambda| \le 1$ and unstable if $|\lambda| > 1$. 

The partition function, $Y$, of $k$-mers on the HL can be obtained, similarly to the recursion relations for the ppf's, by considering all the possible ways of attaching four subtrees to a central square. It can be written as
\begin{equation}
 Y = \sum_{n=0}^{k-1} [g'(n,0)g(k-n-1,0) + g'(0,n)g(0,k-n-1) ] = [g(0,0)]^4 y,
\end{equation}
where $y$ is given by
\begin{equation}
 y = \sum_{n=0}^{k-1} [ f(n,0)R(k-n-1,0) +  f(0,n)R(0,k-n-1) ].
\end{equation}
Then, the average number of $k$-mer bonds reaching the root site (from above) in the $x$ and $y$ directions are
\begin{equation}
 n_x = \frac{\sum_{n=1}^{k-1} f(n,0)R(k-n-1,0)}{y} \quad \text{and} \quad n_y = \frac{\sum_{n=1}^{k-1} f(0,n)R(0,k-n-1)}{y},
\end{equation}
and we may define a nematic order parameter as
\begin{equation}
 \psi = \frac{|n_x - n_y|}{n_x + n_y}.
 \label{eqOrdPar}
\end{equation}

The bulk free energy \textit{per site}, $\phi_b$, at the central square of the Husimi lattice reads \cite{Gujrati,MinosJurgen,tiagoPol}
\begin{equation}
 \phi_b = - \frac{k_B T}{4} \ln\left[ \frac{Y'}{Y^3} \right],
\label{eqFEL1}
 \end{equation}
where $Y$ and $Y'$ denote the partition functions in generations $M$ and $M+1$, respectively. Thereby, it is a easy task to show that
\begin{equation}
 \phi_b = - k_B T \ln\left[ \frac{f(0,0)}{y^{1/2}} \right].
\end{equation}
So, the dimensionless entropy (in units of $k_B$) is given by
\begin{equation}
 s = - \frac{1}{k_B}\left( \frac{\partial \phi_b}{\partial T} \right) = \ln\left[ \frac{f(0,0)}{y^{1/2}} \right].
 \label{eqSRH}
\end{equation}

For dimers, one finds that $g'(1,0)=g'(0,1)$ in Eqs. \ref{eqRRsL1}, so that we can deal with a single ratio $R \equiv [g(1,0)+g(0,1)]/g(0,0)$, whose physical fixed point solution is $R = \sqrt[4]{2}$. Moreover, it is quite easy to verify that $f(0,0) = R(2+R^2)$ and $y=2+4R^2+R^4$ in this case, so that $s = \ln\left[ \frac{\sqrt[4]{2}(2+\sqrt{2})}{2\sqrt{1+\sqrt{2}}} \right] = 0.267399998\ldots$, in agreement with the result from Ref.  \cite{JurgenMario}.  

For larger rods, we solved the fixed point equations \ref{fpe} numerically and then obtained the entropy and nematic order parameter for the fixed points, which are depicted in Tabs. \ref{TabEntDiag} and \ref{TabEntKS}. The dominant eigenvalue of the jacobian of the recursion relations was also determined. For $k<4$, only the isotropic fixed point is found, and $|\lambda^{(I)}|<1$, so that the isotropic fixed point is stable [see Tab. \ref{TabRHlambda}]. For $k \ge 4$, two fixed points are found, the isotopic and the nematic one. The isotropic fixed point is unstable, while the nematic one is stable, as shows Tab. \ref{TabRHlambda}. The entropy for the isotropic fixed point is negative for $k \ge 4$ [see Tabs. \ref{TabEntDiag} and \ref{TabEntKS}], signalling also that this fixed point is not physical for this range of rod sizes on the HL. We notice also that $|\lambda^{(N)}|$, for the nematic phase, rapidly approaches 1 as $k$ increases. This explains why in the direct calculation of the fixed points, by iteration of the recursion relations for the ratios, the convergence becomes slower as the rods grow. Also, as already mentioned, the leading eigenvalues are in general complex, which means that the values of the ratios do not converge uniformly to the fixed point, and sometimes the system may be trapped for many iterations in a sequence of values which resembles a limit cycle. These features are seen also in the trees built with larger cells, discussed in the following sections.

By comparing the entropies in Tabs. \ref{TabEntDiag} and \ref{TabEntKS} for the HL and for the Bethe lattice, we may note that they are always larger in the former case, with the largest increase, of the order of 50\%, being observed for trimers. The differences between both estimates become quite small as $k$ increases, so results for both lattices in the large $k$ limit follow the same asymptotic behavior.  This is indeed expected, once a large rod will not distinguish too much between an underlying Bethe or an ordinary Husimi lattice.  It is very likely that, for this same reason, we are still finding a nematic phase for $k>7$ in the full lattice limit, though the nematic order parameters are smaller in the Husimi lattice results.

\begin{table}
\caption{Results for the leading eigenvalue of the jacobian calculated at the isotropic ($\lambda^{(I)}$) and the nematic ($\lambda^{(N)}$) fixed points. The real and imaginary parts of $\lambda$ are shown into parenthesis followed by the corresponding modulus.}
\label{TabRHlambda}
\begin{ruledtabular}
\begin{tabular}{lllll}
$k$ & $\lambda^{(I)}$ & $|\lambda^{(I)}|$ & $\lambda^{(N)}$ & $|\lambda^{(N)}|$ \\
\hline
2 & (-0.6568542, 0.0000000) & 0.6568542  & - & -\\
3 & (-0.4270509, 0.7251423) &  0.8415485 & - & -\\
4 & (1.3239469, 0.0000000) & 1.3239469 &
(-0.96392875, 0.00000000) & 0.96392875 \\
5 & (1.7462874, 0.0000000) & 1.7462874 &
(-0.80569924, 0.58665104) & 0.99664975 \\
6 & (2.0057972, 0.0000000) & 2.0057972 &
(-0.99974075, 0.00000000) & 0.99974075 \\
7 & (2.1768442, 0.0000000) & 2.1768442 &
(0.62350679, 0.78179613) & 0.99998295 \\
8 & (2.2970438, 0.0000000) & 2.2970438 &
(0.70710773, 0.70710447) & 0.99999904 \\
9 & (2.3859984, 0.0000000) & 2.3859984 &
(0.76604448, 0.64278748) & 0.99999995 \\
10 & (2.4545486, 0.0000000) & 2.4545486 &
(0.80901699, 0.58778524) & 0.99999999 
\end{tabular}
\end{ruledtabular}
\end{table}

\section{Husimi lattices built with diagonal square clusters}
\label{secDiag}

Now, we consider rods defined on HLs whose building blocks (BBs) are diagonal square lattices, with $2L(L+1)$ sites, as shown in Fig. \ref{FigBBs}. The effective lateral size $L$ defines the level of approximation for the square lattice.

\subsection{Preliminaries}

\begin{figure}[b]
\centering
\includegraphics[width=9.0cm]{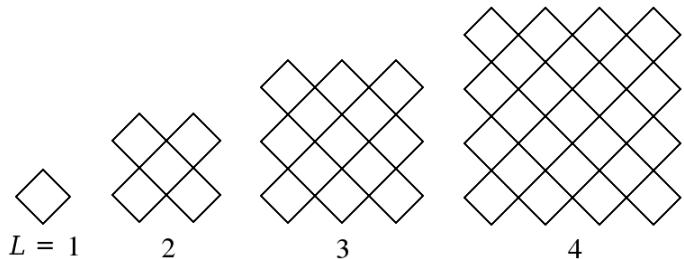}
\caption{Building blocks of generalized HLs for levels up to $L=4$.}
\label{FigBBs}
\end{figure}

\begin{figure}[t]
\centering
\includegraphics[width=8.0cm]{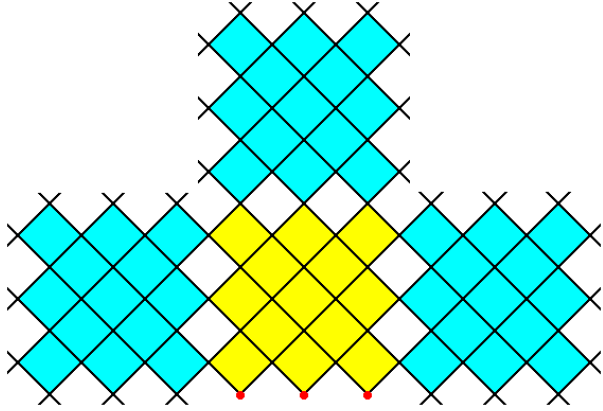}
\caption{Rooted building block with three subtrees attached to it, for $L=3$. Different generations of the tree are indicated by different colors. The three sites of the root line are represented by the red dots.}
\label{FigHL3}
\end{figure}

The solutions of the problem in these generalized HLs follow the very same steps as in the ordinary case ($L=1$) of the previous section, but now two adjacent BBs (in consecutive generations of the tree) are connected by $L$ sites. In Fig. \ref{FigBBs} the BB's for $L$ between 1 and 4 are shown. The ppf's may then be determined through the configurations of $k$-mer bonds reaching (from above) the $L$ root sites, such that we have now to deal with root lines of rooted BBs [see Fig. \ref{FigHL3}]. As above, the configuration of a given root site $s$ can be denoted by $(i_s,j_s)$, with $i_s$ [$j_s$] accounting for the number of incorporated monomers in the incident rod in the $x$ [$y$] direction. Hence, the ppf's for a $L$-level HL are given by $g(i_1,j_1; i_2,j_2;\ldots; i_L,j_L)$, with $i_s,j_s = 0,\ldots,k-1$ for $s=1,\ldots,L$. Although each root site can be found in $2k-1$ states, the total number of configurations for the root line is much smaller than $(2k-1)^L$. In fact, the state of a given root site $s$ can impose restrictions on the configurations of other root sites. For example, if $j_s=n>0$, then, $i_{s+1} = 0$, $i_{s+2} \leqslant 1$, $i_{s+3} \leqslant 2,\ldots$, $i_{s+n} \leqslant n-1$, assuming that $s+n \leqslant L$. So, it becomes very cumbersome to determine all the allowed configurations for the root line ``by hand'' as $L$ and $k$ increases, but this can be done computationally, which is the way we will work hereafter. The numbers $N_{kL}$ of possible configurations for the root line when $k$-mers are placed on a $L$-level HL are displayed in Tab. \ref{TabNconf}. 

We notice that, instead of using the set $\{i_1,j_1; i_2,j_2;\ldots; i_L,j_L\}$ to represent a given configuration, it is very convenient to order and label these configurations by a single integer. In this way, the ppf's can be simply denoted by $g(i)$, with $i=0,\ldots,N_{kL}-1$., where $g(0)$ will always represent the configuration chosen to be placed in the denominator of the ratios of ppf's defined below.

Similarly to the ordinary HL, we may write down a set of recursion relations (RRs) for these ppf's by considering the process of building a ($M+1$)-generation subtree by attaching three $M$-generation ones to a rooted BB, as illustrated in Fig. \ref{FigHL3}, for $L=3$. In general, the RR for the ppf associated with a given configuration $i$ can be written as
\begin{equation}
g'(i) = \sum_{l=0}^{N_{kL}} \sum_{t=0}^{N_{kL}} \sum_{r=0}^{N_{kL}} m_{kL}(i;l,t,r) g(l) g(t) g(r),
\label{eqRRsLqq}
\end{equation}
where the integers $l$, $t$ and $r$ set the configurations of the subtrees attaching respectively at left, top and right side of the rooted BB, whose root line is at configuration $i$. Note that rods from the attaching subtrees, as well as those in the root line may extend to the interior of the rooted BB and may even cross it. Therefore, a large number of combinations of the configurations $(i;l,t,r)$ are forbidden, because they would lead to attrition of rods and/or inconsistencies in their continuity (or lengths). The integers $m_{kL}(i;l,t,r)$ account for this in the RRs, vanishing in these cases. Moreover, once an allowed set of configurations is found, since it already determines the occupancy of some (or all) bulk sites, we have to look for the possible free sites. If they do not exist, $m_{kL}(i;l,t,r)=1$; otherwise, $m_{kL}(i;l,t,r)$ will be the number of ways of covering them with $k$-mers. For many sets $(i;l,t,r)$, specially for $k \gtrsim L$, a full coverage is not possible, so that $m_{kL}(i;l,t,r)=0$. Hence, the sum over $N_{kL}^3$ terms in Eq. \ref{eqRRsLqq} has actually a much smaller number of non-null contributions. As an example, for $L=4$ and $k=4$, one has only $11079$ non-null terms for $g'(0)$, whereas $N_{kL}^3 \approx 5.6 \times 10^{8}$.

\begin{table}
\caption{Number of configurations $N_{k,L}$ for the root line of a diagonal-square HL of level $L$, with $k$-mers placed on it.}
\label{TabNconf}
\begin{ruledtabular}
\begin{tabular}{rrrrrrrr}
$L\setminus k$ &    $2$  &     3  &      4  &     5  &      6  &       7    \\
\hline
2  &    8  &    20  &     36  &    56  &     80  &     108    \\
3  &   21  &    77  &    175  &   325  &    539  &     829    \\
4  &   55  &   292  &    826  &  1820  &   3498  &    6136    \\
5  &  144  &  1098  &   3828  &  9956  &  22184  &     --     \\
\end{tabular}
\end{ruledtabular}
\end{table}

\begin{figure}[t]
\centering
\includegraphics[width=8.0cm]{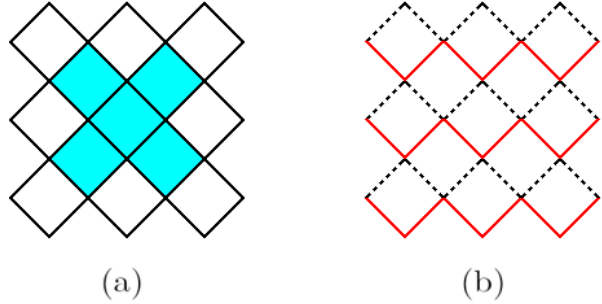}
\caption{Building blocks, for $L=3$. In (a) the colored part highlights the BB for $L-1$. The definition of the ``up'' (solid red) and ``down'' (dotted black) zigzag lines forming the BB is presented in (b).}
\label{FigLL1}
\end{figure}

It is clear from Eq. \ref{eqRRsLqq} that, once we known the possible configurations for the root line, we only need to determine the variable $m_{kL}(i;l,t,r)$ to have the RRs. At first, this can be (computationally) done by fixing the configurations $(i;l,t,r)$ in the four sides of the rooted BB (RBB) and, then, checking for attritions and discontinuities in length. If they are found, one makes $m_{kL}(i;l,t,r)=0$ and goes to the next set of configurations. Otherwise, there are some options to determine the bulk configurations. For instance, we may use, e.g., the Hoshen-Kopelman algorithm \cite{hk76} to identify the possible clusters of empty sites in the interior of the RBB and, then, try to fully cover these clusters (if they exist) with $k$-mers through an exact enumeration process to find $m_{kL}(i;l,t,r)$. It turns out however that this complicated procedure can become very computationally demanding already for relatively small $L$'s; at one hand, because $N_{kL}$ becomes large for large $k$ and, on the other hand, because there are much bulk configurations for small $k$. Another possibility is the use of the RRs for the case $L-1$ to obtain those for $L$, as recently done for hard squares in \cite{Nathann21}. In fact, as illustrated in Fig. \ref{FigLL1}(a), the central portion of a $L$-level RBB can be seen as the RBB for $L-1$. Thereby, for each configuration $(i;l,t,r)$ of the three incoming subtrees ($l,t,r$) for a given root line $i$ of the $L$-level system, we can run over all the allowed configurations for the $L-1$ case, looking for those that fit at the center of the larger RBB (satisfying the full occupancy condition and etc). We find in this implementation that, since we have to compare all the allowed configurations at four sides of the $L$-level RBB with all the ones for $L-1$, this becomes slow already for not so large $L$ and $k$.

Therefore, we use a different strategy to obtain the variable $m_{kL}(i;l,t,r)$ and so the RRs. For given $L$ and $k$, beyond determining the configurations of the zigzag root line (let us refer to it as a line of type ``up''), we determine also the configurations for such line flipped upside down (the ``down'' line). As shown in Fig. \ref{FigLL1}(b), a $L$-level RBB can be built by alternately adding $L$ ``down'' lines over $L$ ``up'' ones and vice-versa. So, with the line configurations at hand, we construct two transfer matrices (for open boundary conditions): one for lines ``down'' adding over ``up'' ones and another one for the opposite case. Using these matrices, it is quite simple to build up all the possible rod configurations for the RBB for a fixed root line configuration. The process of obtaining the RRs for level $L$ (for a given $k$) becomes even more optimized if one starts with a line of type ``down'' of size $L+1$ and alternately adds $L$ ``up'' and $L-1$ ``down'' lines over it, all of them for $L+1$. This automatically furnishes the configurations of the RBB, as well as those of the incoming subtrees at its left and right sides. Note that, thanks to the transfer matrices, we only visit allowed configurations along this process, what turns it much more effective than the approaches discussed above.

As before, we work with ratios of ppf's, which will be defined in general as $R(i) = g(i)/g(0)$, for $i=1,\ldots,N_{kL}-1$. At first, the ppf chosen to be in denominator [and generically labeled here as $g(0)$] can be any of the $N_{kL}$ ones and, in most cases, $g(0)$ will represent the configuration where no rods reach the root sites from above, as in the previous section. For some few particular cases (e.g., $L=2$ and $k=4$, and $L=4$ and $k=4$) this choice for $g(0)$ yields divergent ratios. Hence, in such situations, $g(0)$ will represent other configuration, different from $(0,0;0,0;\ldots;0,0)$, which leads to convergent ratios in the thermodynamic limit. In any case, the RRs for the ppf's can always be written as $g'(i)=g(0)^3 f(i)$, with $f(i)$ being a sum depending only on the ratios. Thereby, the RRs for the ratios take the form $R'(i) = f(i)/f(0)$.

By connecting four subtrees to a central BB [summing over all the possible ways of doing this, satisfying the full occupancy, avoiding attritions and etc.] we obtain the partition function, $Y$, which may be written, in general, as
\begin{equation}
 Y = \sum_{i=0}^{N_{kL}}\sum_{j=0}^{N_{kL}} \Delta_{ij} g(i)g'(j) = g(0)^4 y,
\end{equation}
where $\Delta_{ij}=1$ if the configurations $i$ and $j$ match at the root line; and $\Delta_{ij}=0$ otherwise. In addition, 
\begin{equation}
 y = \sum_{i=0}^{N_{kL}}\sum_{j=0}^{N_{kL}} \Delta_{ij} R(i)f(j)
\end{equation}
only depends on the ratios. Therefore, the average number of $k$-mer bonds reaching the root line (from above) in the $x$ and $y$ directions read
\begin{equation}
 n_{x} = \frac{\sum_{i=0}^{N_{kL}} \sum_{j=0}^{N_{kL}} \Delta_{ij} \gamma^{(x)}_j R(i)f(j)}{y} \quad \text{and} \quad n_{y} = \frac{\sum_{i=0}^{N_{kL}}\sum_{j=0}^{N_{kL}} \Delta_{ij} \gamma^{(y)}_j R(i)f(j)}{y}.
 \label{eqDens}
\end{equation}
Here, $\gamma_j^{(s)}$ is the number of bonds in the root line at configuration $j$ in the direction $s=x,y$. Then, the nematic order parameter can be calculated from Eq. \ref{eqOrdPar}.

For $L$-level HLs, the bulk free energy per site in Eq. \ref{eqFEL1} trivially generalizes to \cite{Nathann21}
\begin{equation}
 \phi_b = - \frac{k_B T}{2 V_{eff}} \ln\left[ \frac{Y'}{Y^3} \right],
\label{eqFELqq}
 \end{equation}
with $V_{eff}=2L^2$ being the effective number of sites in each BB, once the $4L$ sites shared between two generations of the tree contribute as $2L$. This leads to the dimensionless entropy
\begin{equation}
 s_L(k) = - \frac{1}{k_B}\left( \frac{\partial \phi_b}{\partial T} \right) = \frac{1}{L^2}\ln\left[ \frac{f(0)}{y^{1/2}} \right].
 \label{eqEntLqq}
\end{equation}

As it will be seen in what follows, for some particular values of $L$ and $k$, the RRs converge to a limit cycle of period 2, instead of a fixed point, so that $R'_B(i)=f_A(i)/f_A(0)$ and $R'_A(i)=f_B(i)/f_B(0)$, with $A$ and $B$ denoting the different points of the cycle. In this case, the convergent part of the partition function might also oscillate between two values ($y_{A}$ and $y_{B}$) and, as demonstrated in the Appendix, a more appropriate definition for $\phi_b$ is
\begin{equation}
 \phi_b = - \frac{k_B T}{8 V_{eff}} \ln\left[ \frac{Y''}{Y^9} \right]
\label{eqFELqqc2}
 \end{equation}
where $Y''$ ($Y$) is the partition function for generation $M+2$ ($M$). Then, we obtain the dimensionless entropies [see the Appendix]
\begin{equation}
 s_{L,A}(k) = \frac{1}{4 L^2}\ln\left[ \frac{f_A(0)^3 f_B(0)}{y_{A}^{2}} \right], \quad \text{and} \quad s_{L,B}(k) = \frac{1}{4 L^2}\ln\left[ \frac{f_A(0) f_B(0)^3}{y_{B}^{2}} \right].
\label{eqEntLqqc2}
 \end{equation}
As expected, in the case of a fixed point, where $f_A(0)=f_B(0)=f(0)$ and $y_A = y_B=y$, these generalized definitions reduce to Eq. \ref{eqEntLqq}. 

For all $L$ and $k$ analyzed here, we find $s_{L,A}(k) = s_{L,B}(k)$, so that the indexes $A$ and $B$ will be suppressed from the entropies below. On the other hand, $n_{x,A}$ and $n_{y,A}$ (calculated from Eq. \ref{eqDens} with $R_A$, $f_A$ and $y_A$) are different from $n_{x,B}$ and $n_{y,B}$, yielding different values for $\psi_A$ and $\psi_B$, calculated from Eq. \ref{eqOrdPar}. Hence, in this case, the order parameter presented below is $\psi = (\psi_A + \psi_B)/2$.

\subsection{Results}

The entropies obtained for the diagonal square HLs are summarized in Tab. \ref{TabEntDiag}, along with those for the ordinary HL and the Bethe lattice, where the values of the order parameter for the nematic phase are also shown. The few situations where the RRs converge to limit cycles are also indicated, with fixed points being find in the rest. We remark that two types of cycles of period 2 are found: regular ones, for which the RRs alternate between two sets of finite values; and ``diverging" ones, where some ratios converge to finite values, but others oscillate between diverging and vanishing values. In this case, there is no suitable choice for the denominator of the RRs to prevent the divergences. Namely, by changing the configuration in the denominator, we simply change the sets of finite, diverging and vanishing RRs, while the finite entropy remains the same.

\begin{table}[h]
\caption{Entropy $s_L(k)$, for diagonal-square HLs of levels $L \leq 5$ and several $k$'s, for the isotropic ($I$) and nematic ($N$) phases. The bottom line presents the extrapolated values $s_{\infty}^{(I)}$, obtained from 3-pt extrapolations of the set ($s_1^{(I)},s_3^{(I)},s_5^{(I)}$) to $L \rightarrow \infty$. The values of the nematic order parameters $\psi$ (for the nematic phase) are shown between parentheses.}
\label{TabEntDiag}
\begin{ruledtabular}
\begin{tabular}{ccccccc}
$k$        &     2         &      3        &       4                   &       5                   &       6           &       7            \\
\hline
$s_0^{(I)}$&    0.2616241  &    0.0566330  &    -6.764415E-02          &   -0.1524737              &    -0.2146781     &   -0.2625527        \\ 
$s_0^{(N)}$&    --         &    --         &     4.276367E-03          &    3.247911E-04           &     2.147242E-05  &    1.2144921E-06     \\
$(\psi)$   &               &               &     (0.962250)            &    (0.996702)             &     (0.999741)    &    (0.999983)       \\ 
\hline
$s_1^{(I)}$&    0.2673999  &    0.0827527  &    -1.771135E-02          &   -8.037280E-02           &    -0.1230766     &   -0.1540139        \\ 
$s_1^{(N)}$&    --         &    --         &     5.386386E-03          &    3.394689E-04           &     2.161841E-05  &    1.215430E-06     \\
$(\psi)$   &               &               &     (0.936837)            &    (0.996391)             &     (0.999738)    &    (0.999982)       \\ 
\hline
$s_2^{(I)}$&    0.2822379  &    0.1215620  &     6.938599E-02          &    6.063561E-03           &     2.068819E-02  &    -2.970555E-02               \\
$s_2^{(N)}$&    --         &    --         &     6.947659E-02          &    8.194540E-03           &     --            &    2.048538E-04     \\
$(\psi)$   &               &               &    (7.594707E-02)         &    (0.740571)             &                   &    (0.995860)       \\
\hline
$s_3^{(I)}$&    0.2854815  &    0.1463681  &     6.837276E-02$^{(a)}$  &    3.793798E-02$^{(b)}$   &     3.099252E-02  &    -4.709141E-03              \\
$s_3^{(N)}$&    --         &    --         &     --                    &    --                     &     3.123758E-02  &    1.582462E-03     \\
$(\psi)$   &               &               &                           &                           &    (9.879675E-02) &    (0.939622)       \\
\hline
$s_4^{(I)}$&    0.2878447  &    0.1455480  &     9.095362E-02          &    4.313372E-02$^{(b)}$   &     3.299486E-02$^{(b)}$  &   1.831432E-02$^{(b)}$               \\
$s_4^{(N)}$&    --         &    --         &     --                    &    --                     &     --            &           1.832327E-02$^{(b)}$               \\
$(\psi)$   &               &               &                           &                           &                   &           (6.378773E-02)          \\
\hline
$s_5^{(I)}$&    0.2887289  &    0.1487062  &     8.763683E-02          &    6.390494E-02            &     3.2731535E-02$^{(a)}$ &    --               \\
$s_5^{(N)}$&    --         &    --         &     8.765191E-02          &    --                      &     --            &    --               \\
$(\psi)$   &               &               &     (7.50218E-02)         &                            &                   &                     \\
\hline
$s_{\infty}^{(I)}$&  0.29211  &  0.14930  &  0.11476  &  0.09943  &  0.03294  & --  \\
\end{tabular}
\end{ruledtabular}
{}$^{(a)}$ Divergent limit cycle-2.\\
{}$^{(b)}$ Convergent limit cycle-2.
\end{table}

\begin{figure}[t]
\centering
\includegraphics[width=10.0cm]{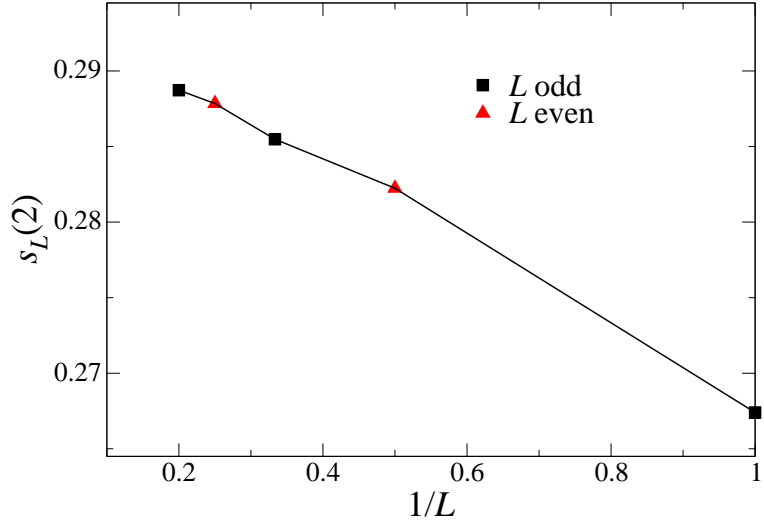}
\caption{Entropy for $k=2$ against $1/L$, with $L=1,2,...,5$. Black filled squares represents odd values of $L$, while the red triangles even values.}
\label{FigEntk2}
\end{figure}

For dimers and trimers, only the isotropic phase is found in the system, in agreement with the results for the Bethe (and RLTL) lattice \cite{drs11} and in consonance with the expected behavior for the square lattice \cite{gd07}. Although $s_L(2)$ increases monotonically with $L$, its convergence depends on the $L$-parity, as shown in Fig. \ref{FigEntk2}. This is even more clear for $k=3$, once $s_4(3)<s_3(3)$ in Tab. \ref{TabEntDiag} and happens also for larger rods. This parity effect hampers the data extrapolation for $L \rightarrow \infty$ (i.e, for the infinite square lattice limit), once we need to analyze high levels to end with few points to extrapolate. For instance, to perform five-point (5-pt) extrapolations of the entropies for odd $L$, assuming, e.g., power-law corrections of the form
\begin{equation}
s_L = s_{\infty} + a_1 \mathcal{L}^{-\alpha_1} + a_2 \mathcal{L}^{-\alpha_2}
\label{eqFSS}
\end{equation}
(with $\mathcal{L}$ being related to $L$ as defined just below), we should have $s_L$, at least, for $L \leq 9$ to extrapolate the set ($s_1$, $s_3$,...,$s_9$), but this is unfeasible. In fact, our results are restricted to $L \le 5$, once $s_5(k)$ is already hard of obtaining, specially for large $k$, due to the fast increase of $N_{k,L}$ with both $L$ and $k$ [see Tab. \ref{TabNconf}]. This is the reason for the absence of results for $L=5$ and $k=7$ in Tabs. \ref{TabNconf} and \ref{TabEntDiag}. Hence, we can perform only 3-pt extrapolations, assuming that $a_2=0$ in Eq. \ref{eqFSS}, which has thus three unknowns: $s_{\infty}$, $a_1$ and  $\alpha_1$. If we extrapolate considering $s_0(k)$ in the set of even $L$'s [i.e., $(s_0,s_2,s_4)$], we obtain $s_{\infty}(2) = 0.30056$ and $s_{\infty}(3) = 0.27148$, when $\mathcal{L}=L^*$, with $L^*$ being the square root of the \textit{total} number of sites in each BB (such that $L^*=1$ for the BL and $L^*=\sqrt{2L(L+1)}$ for the HLs). These entropies differ, respectively, by 3\% and 71\% from the expected results for dimers and trimers on the square lattice. We notice that these deviations increases if one defines $\mathcal{L}$ as the square root of the \textit{effective} number of sites in each BB (i.e., $\mathcal{L}=1/2$ for the BL and $\mathcal{L}=\sqrt{2}L$ for the HLs), so that we will always use $\mathcal{L}=L^*$ in the discussion below. The inaccuracy in these estimates, particularly in $s_{\infty}(3)$, certainly happens because the BL solution is still a very crude approximation for the rods' behavior on the square lattice. In fact, by extrapolating the set $(s_1,s_3,s_5)$ we obtain much better results, which are depicted in Tab. \ref{TabEntDiag} as $s_{\infty}^{(I)}$, deviating by $\approx 0.2$\% from the exact entropy for the square lattice in the case of dimers, while for trimers one finds a difference of $\approx 6$\% from the value estimated in Ref. \cite{gdj07}. This demonstrates that, by increasing the level $L$, the extrapolated entropies get closer to the square lattice values. Moreover, the variation observed in the deviations suggests that to obtain $s_{\infty}(k)$ with similar accuracy for different $k$'s, we should extrapolate data for levels $L$ that increase with $k$, which is unfortunately not possible. 

Despite this, reasonable estimates are obtained from 3-pt extrapolations of the entropies of the isotropic phase, considering the set of odd-$L$'s, for $k=4$ and $k=6$ [see Tab. \ref{TabEntDiag}], once both $s_{\infty}^{(I)}(4)$ and $s_{\infty}^{(I)}(6)$ are within the intervals determined in Ref. \cite{gp79} for these entropies in the square lattice. For $k=5$, on the other hand, our extrapolated value is out of the range determined by the lower and upper bounds from Ref. \cite{gp79} for the square lattice: $0.04665 \le s(5) \le 0.08805$. A similar issue is observed for $k=7$, in the extrapolations of the set $(s_0,s_2,s_4)$ for both the isotropic and nematic phase.

Although these extrapolations are returning unreliable values in some cases, the results in Tab. \ref{TabEntDiag} are consistent with an isotropic phase in the square lattice ($L \rightarrow \infty$) limit, as expected at full packing \cite{gd07}. For instance, the entropy of this phase becomes less negative as $L$ increases and, with exception of $k=7$ (where it is still negative up to $L=3$), it becomes positive already for $L \ge 2$ in the other cases. Note also that for the higher $L$'s there are several cases where results for the nematic phase are lacking in Tab. \ref{TabEntDiag} and this happens because its fixed point is not found by iterating the RRs. In fact, in such cases, even if one starts the iteration with initial conditions that would yield the symmetry breaking of the nematic phase, the RRs converge to the isotropic fixed point. Although the appearance of the nematic phase becomes rare at higher levels, whenever it shows up, it has an entropy larger than the one for the isotropic phase, for given $k$ and $L$. Therefore, at least when it appears, the nematic phase is the stable one. This is confirmed also by the leading eigenvalue, $\lambda$, of the jacobian matrix, since one finds $|\lambda^{(N)}|<1$ and $|\lambda^{(I)}|>1$ when both phases are present. On the other hand, when only the isotropic phase is found, we obtain $|\lambda^{(I)}|<1$, demonstrating that it is stable in such situations. Despite this disappearance and re-appearance of nematic phase for a given $k$, without any clear rule, when it appears its order parameter is a decreasing function of $L$. Moreover, for higher $L$'s, $s_L^{(I)}(k)$ and $s_L^{(N)}(k)$ are quite close and the difference between them decreases as $L$ augments. As an example, for $k=4$, this difference is $\approx 0.13$\% for $L=2$ and $\approx 0.02$\% for $L=5$. All these results strongly indicate that only the isotropic phase shall exist for $L \rightarrow \infty$.

\section{Generalized Husimi lattices built with regular square clusters}
\label{secKS}

\subsection{Preliminaries}

\begin{figure}[b]
\centering
\includegraphics[width=9.0cm]{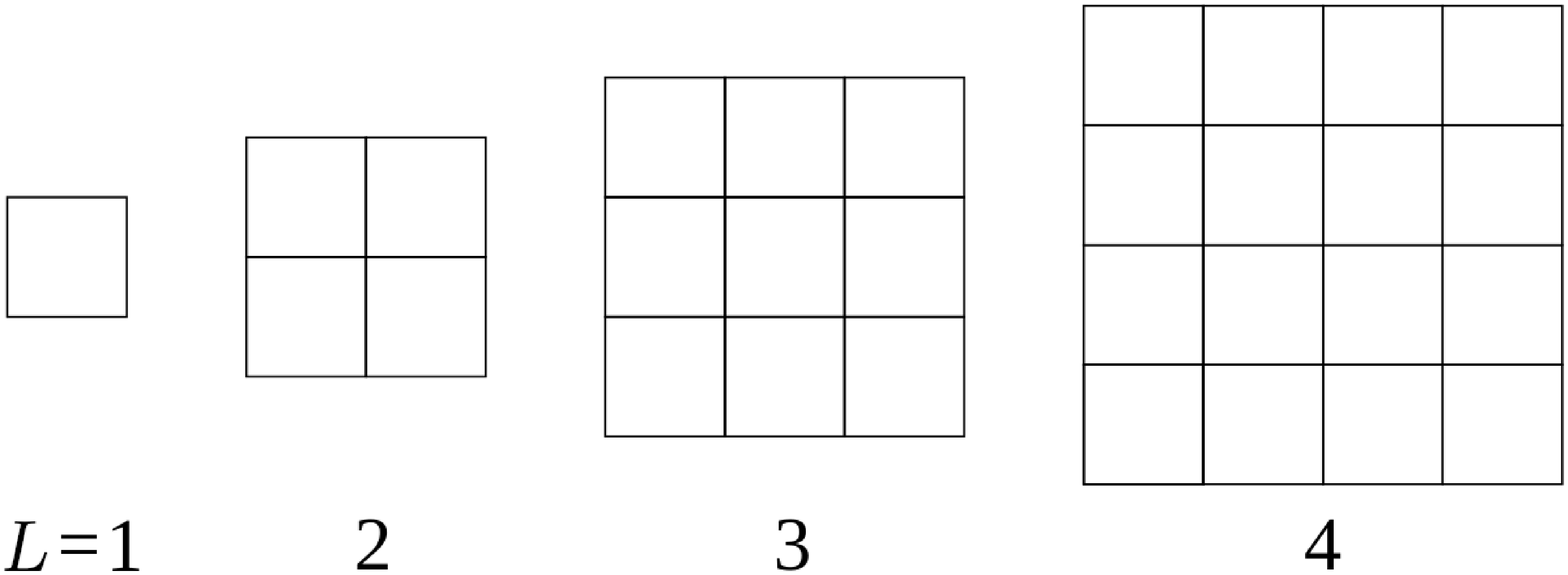}
\caption{Building blocks of Kobaiashi-Suzuki generalized HLs for levels up to $L=4$}
\label{BB-KS}
\end{figure}

\begin{figure}
\centering
\includegraphics[width=9.0cm]{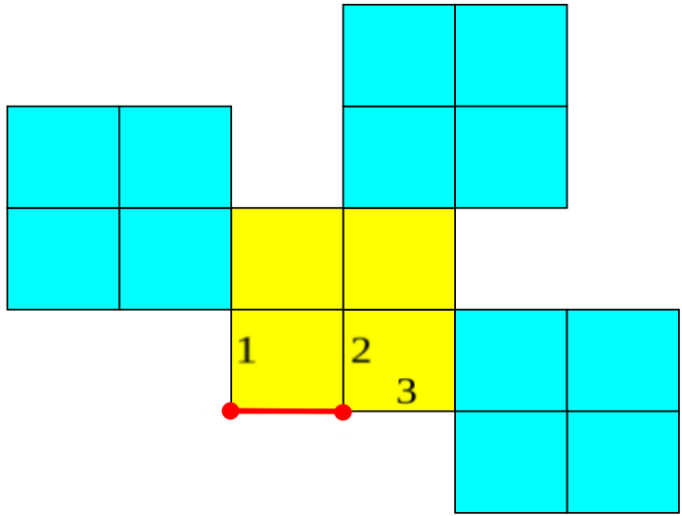}
\caption{Two generations of a $L=2$ subtree, indicated by different colors. The two sites and the edge between them indicated in red at the root building block will be shared with the new block of the next generation. The numbers are the order of the edges incident at the root, which will be used to specify its configuration.}
\label{RR-KS}
\end{figure}

In view of the limitation (to low levels) of the results obtained for the diagonal approximation in the previous section, we will analyze the rods in another sequence of treelike lattices built with growing clusters, which was proposed by Kobaiashi and Suzuki in 1993 \cite{ks93}. In this generalization of the Husimi lattice, the building blocks are regular square lattices with lateral size $L+1$. Some examples of them, for $L \le 4$, are shown in Fig. \ref{BB-KS}. One key difference of this sequence relative to the one considered in the previous section is that adjacent building blocks, of successive generations of the tree, share $L-1$ edges also, besides the $L$ sites (see Fig. \ref{RR-KS}).

As before, we proceed writing down recursion relations for the ppf's. We will describe in some detail the solution for the case of dimers ($k=2$) on the $L=2$ lattice. In Fig. \ref{RR-KS} a subtree is shown. One point which has to be noticed is that since for sufficiently large rods nematic order is expected, the two possible orientations have to be distinguished in the calculations. Thus, we will define two sets of ppf's: $g(i)$ for subtrees whose root edge is in the $x$ direction and $h(i)$ if it is in the $y$ direction.

The configuration of the root sites of a subtree will be defined by the rods which reach them whose starting monomer is located in building blocks at the root or in earlier generations. We notice that there are $L+1$ edges incident on the root sites, and we may label the configuration specifying the number of monomers already incorporated into the rod, so that these numbers will be in the range $[0,k-1]$. There is a constraint for the last pair of numbers, since they correspond to two edges which reach the same root site: at least one of them has to be equal to zero. Also, if an endpoint monomer reaches the rightmost site of the root, the cases where this monomer belongs to an horizontal or vertical rod are equivalent and are included in the same configuration. Therefore, the number of configurations will be:
\begin{subequations}
\begin{eqnarray}
N_{k,L}&=&k^{L-k+1}\frac{k^k-1}{k-1}-k^{L-1},\,\,\mbox{for } k \le L+1;\\
N_{k,L}&=&\frac{k^{L+1}-1}{k-1}+k-L-1-k^{L-1},\,\,\mbox{for } k \ge L+1.
\end{eqnarray}
\label{nc}
\end{subequations}
These numbers are explicitly shown in Tab. \ref{TabNconf-KS} for the parameters ($k$ and $L$) analyzed here. It is noteworthy that these $N_{k,L}$ are much smaller than those in Tab. \ref{TabNconf} for the diagonal case. This will allow us to investigate higher levels here, at least for the smaller $k$-mers.

\begin{table}[b]
\caption{Number of configurations $N_{k,L}$ for the root line of a K-S HL of level $L$, with $k$-mers placed on it.}
\label{TabNconf-KS}
\begin{ruledtabular}
\begin{tabular}{rrrrrrrrrr}
$L\setminus k$ &    $2$  &     3  &      4  &     5  &      6  &       7  \\
\hline
2&      4&     10&     18&     28&     40&     54 \\   
3&      8&     30&     69&    132&    225&    354 \\  
4&     16&     90&    276&    656&   1340&   2460 \\  
5&     32&    270&   1104&   3280&   8035&  17208 \\  
6&     64&    810&   4416&  16400&  48210& 120450 \\  
7&    128&   2430&  17664&	82000&	289260&	843150\\  
\end{tabular}
\end{ruledtabular}
\end{table}

We will adopt a particular order of the edges which are incident on the root sites, starting at the corner of the root building block, and moving along the edge of it. In the example for $L=2$ depicted in Fig. \ref{RR-KS} this order is shown. We may associate an integer $i_s$ to each state, starting with $i_s=0$ for the state $(0,0,\ldots,0)$ and $i_s=N_{k,L}-1$ for the last state $(k-1,k-1,\ldots,0,k-1)$. For example, in the case of dimers ($k=2$) on the tree with $L=2$, the number of states is $N_{2,2}=4$, and the configurations associated to these states  are $i_s=0$: $(0,0,0)$; $i_s=1$: $(1,0,0)$; $i_s=2$: $(0,1,0)$ and $(0,0,1)$; $i_s=3$: $(1,1,0)$ and $(1,0,1)$. 

The procedure to obtain the coefficients of the recursion relations for the ppf's is similar to the one employed in the case of diagonal building blocks, so we will not discuss it in detail here. In general, the recursion relations are given by:
\begin{subequations}
    \begin{eqnarray}
      g^\prime(i)&=&\sum_{l=0}^{N_{k,L}-1} \sum_{t=0}^{N_{k,L}-1} \sum_{r=0}^{N_{k,L}-1} m_{k,L}(i;l,t,r) h(l) g(t) h(r),\\
      h^\prime(i)&=&\sum_{l=0}^{N_{k,L}-1} \sum_{t=0}^{N_{k,L}-1} \sum_{r=0}^{N_{k,L}-1} m_{k,L}(i;l,t,r) g(l) h(t) g(r).
    \end{eqnarray}
    \label{rrksgh}
\end{subequations}
Defining the ratios of the ppf's $R(i)=g(i)/g(0)$ and $S(i)=h(i)/h(0)$, we may write the recursion relations for the ratios of ppf's, which are of the form $R^\prime(i)=f(i)/f(0)$ and $S^\prime(i)=f^*(i)/f^*(0)$ with
\begin{subequations}
  \begin{eqnarray}
      f(i)&=&\sum_{l=0}^{N_{k,L}-1} \sum_{t=0}^{N_{k,L}-1} \sum_{r=0}^{N_{k,L}-1} m_{k,L}(i;l,t,r) S(l) R(t) S(r),\\
      f^*(i)&=&\sum_{l=0}^{N_{k,L}-1} \sum_{t=0}^{N_{k,L}-1} \sum_{r=0}^{N_{k,L}-1} m_{k,L}(i;l,t,r) R(l) S(t) R(r).
    \end{eqnarray}
    \label{rrksrs}
\end{subequations}
We notice that, by definition, $R(0)=S(0)=1$ and that the recursion relations may be rewritten as $g^\prime(i)=g(0)h(0)^2f(i)$ and $h^\prime(i)=g(0)^2h(0)f^*(i)$.

As usual, to obtain the partition function of the rods on the tree, we connect four subtrees to the central building block, leading to:
\begin{equation}
Y=\sum_{i=0}^{N_{k,L}-1}\sum_{j=0}^{N_{k,L}-1}\Delta_{i,j}g(i)g^\prime(j)=g(0)^2h(0)^2 y,
\end{equation}
where 
\begin{equation}
y=\sum_{i=0}^{N_{k,L}-1}\sum_{j=0}^{N_{k,L}-1}\Delta_{i,j} R(i)f(j)
\end{equation}
is a polynomial in the ratios also. 
The bulk entropy per site in the present case, when the RRs converge to a fixed point, is given by:
\begin{equation}
s_{L}(k)=\frac{1}{2V_{eff}}\ln\left(\frac{Y^\prime}{Y^3}\right)=\frac{1}{L^2+1}\ln\left(\frac{f(0)f^*(0)}{y}\right),
\end{equation}
where the effective number of sites of a building block is $V_{eff}=L^2+1$. 

Similarly to the diagonal-square HL of the previous section, we also find the RRs converging to limit cycles of period 2 here and, in such cases, the appropriate definitions for the entropy are (see the Appendix)
\begin{equation}
 s_{L,A}(k)= \frac{1}{4 (L^2+1)}\left[ \frac{f_A(0)^3 f_A^*(0)^3 f_B(0) f_B^*(0)}{y_{A}^4} \right]
\end{equation}
and
\begin{equation}
 s_{L,B}(k)= \frac{1}{4 (L^2+1)}\left[ \frac{f_A(0) f_A^*(0) f_B(0)^3 f_B^*(0)^3}{y_{B}^4} \right].
\end{equation}
As before, we always find $s_{L,A}=s_{L,B}=s_L$ here.

\subsection{Results}

The entropies for the KS-HL are depicted in Tab. \ref{TabEntKS}, along with the values of the order parameter for the nematic phase. The general picture is quite similar to the one for diagonal-square HLs. For example, for dimers and trimers only the isotropic phase is found and $s_{L}^{(I)}(k)>0$ already for the Bethe lattice ($L=0$). For larger rods on low-level HLs, $s_{L}^{(I)}(k)<0$ and the nematic phase is stable [having $s_{L}^{(N)}(k)>0$], but $s_{L}^{(I)}(k)$ becomes less negative as $L$ increases and at some level (which increases with $k$) it becomes positive. At such point the nematic phase stops appearing, though it can eventually show up again, as for $k=5$ for $L=6$. Whenever the nematic phase is found it is the stable one, having the largest entropy and the leading eigenvalue of the jacobian matrix $|\lambda^{(N)}| \le 1$, while $|\lambda^{(I)}| > 1$. When it is absent, however, one finds $|\lambda^{(I)}| \le 1$. Importantly, for a given $k$, the nematic order parameter decreases with $L$. Overall, these results suggest that in the square lattice limit only the isotropic phase shall be observed.

\begin{table}[!h]
\caption{Entropy $s_L(k)$, for KS-HLs of levels $L \leq 5$ and several $k$'s, for the isotropic ($I$) and nematic ($N$) phases. The value of the nematic order parameter (for the nematic phase) is shown into parentesis.}
\label{TabEntKS}
\begin{ruledtabular}
\begin{tabular}{ccccccc}
$k$        &     2         &      3        &       4                   &       5                   &       6           &       7            \\
\hline
$s_0^{(I)}$&    0.2616241  &    0.0566330  &    -6.764415E-02          &   -0.1524737              &    -0.2146781     &   -0.2625527        \\ 
$s_0^{(N)}$&    --         &    --         &     4.276367E-03          &    3.247911E-04           &     2.147242E-05  &    1.2144921E-06     \\
$(\psi)$   &               &               &     (0.962250)            &    (0.996702)             &     (0.999741)    &    (0.999983)       \\ 
\hline
$s_1^{(I)}$&    0.2673999  &    0.0827527  &    -1.771135E-02          &   -8.037280E-02           &    -0.1230766     &   -0.1540139        \\ 
$s_1^{(N)}$&    --         &    --         &     5.386386E-03          &    3.394689E-04           &     2.161841E-05  &    1.215430E-06     \\
$(\Psi)$   &               &               &     (0.936837)            &    (0.996391)             &     (0.999738)    &    (0.999982)       \\ 
\hline
$s_2^{(I)}$&    0.2772131  &    0.1086632  &     1.987719E-02          &   -2.396047E-02           &    -5.707920E-02  &    --               \\ 
$s_2^{(N)}$&    --         &    --         &     --                    &    5.994109E-04           &     2.644304E-05  &    1.273296E-06     \\
$(\psi)$   &               &               &                           &    (0.989434)             &     (0.999613)    &    (0.999981)       \\ 
\hline
$s_3^{(I)}$&    0.2822987  &    0.1291433  &     4.912690E-02          &    5.613078E-03           &    -2.419980E-02  &    -3.586687E-02    \\
$s_3^{(N)}$&    --         &    --         &                           &                           &     3.964852E-04  &     2.339824E-06    \\
$(\psi)$   &               &               &                           &                           &     (0.998644)    &     (0.999943)      \\
\hline
$s_4^{(I)}$&    0.2853171  &    0.1372838  &     7.038655E-02          &    2.507253E-02           &    -1.931040E-03  &    -1.785139E-02   \\
$s_4^{(N)}$&    --         &    --         &     --                    &    --                     &     1.305415E-04  &     4.071709E-06   \\
$(\psi)$   &               &               &                           &                           &     (0.992452)    &     (0.999904)     \\
\hline
$s_5^{(I)}$&    0.2869057  &    0.1424772  &     7.740926E-02          &    4.341118E-02           &    1.974750E-02   &                     \\
$s_5^{(N)}$&    --         &    --         &     --                    &    --                     &     --            &                     \\
$(\psi)$   &               &               &                           &                           &                   &                     \\
\hline
$s_6^{(I)}$&    0.2880401  &    0.1470544  &     8.153816E-02          &    4.895803E-02           &                   &                     \\
$s_6^{(N)}$&    --         &    --         &     --                    &    5.009461E-02           &                   &                     \\
$(\psi)$   &               &               &                           &    (0.632976)             &                   &                     \\
\hline
$s_7^{(I)}$&    0.2887282  &    0.1490338  &                           &                           &                   &                     \\
$s_7^{(N)}$&    --         &    --         &     --                    &                           &                   &                     \\
$(\psi)$   &               &               &                           &                           &                   &                     \\
\end{tabular}
\end{ruledtabular}
\end{table}

Although we are obtaining results for higher levels here when compared with the diagonal square HLs, the number of points to extrapolate (for $L \rightarrow \infty$) is still very limited, due to the unfortunate $L$-parity effect. So, as before, we are not able to perform detailed extrapolations considering high-order corrections here. Assuming the finite-size scaling of Eq. \ref{eqFSS} with $a_2=0$, once again, the most accurate results for $k=2$ were obtained by defining $\mathcal{L}$ as the square root of the total number of sites in each BB, which in present case is $\mathcal{L}=L+1$. In fact, this yields $s_{\infty}(2)=0.29483$, $s_{\infty}(2)=0.29200$ and $s_{\infty}(2)=0.29171$ from 3-pt extrapolations of the set $(s_{l-2},s_l,s_{l+2})$ for $l=3$, $4$ and $5$, respectively. A further 3-pt extrapolation of these extrapolated values (for $l \rightarrow \infty$) returns $s_{\infty}(2) = 0.29164$, differing by only $0.03$\% from the exact entropy of dimers on the square lattice. Similar 3-pt extrapolations for trimers give $s_{\infty}(3)=0.16255$ for $l=3$, $s_{\infty}(3)=0.16171$ for $l=4$ and $s_{\infty}(3)=0.16768$ for $l=5$. This fluctuating behavior does not allow us to extrapolate these data for $l \rightarrow \infty$. Anyhow, this indicates that $s_{\infty}(3)=0.164(2)$, which deviates by $\approx 3$\% from the transfer matrix estimate of Ref. \cite{gd07}. We remark that 4-pt extrapolations assuming the existence of logarithmic corrections [in the form $s_L = s_{\infty} + a L^{-\alpha} \log(L)^{\beta}$ or $s_L = s_{\infty} + a L^{-\alpha} \log(b L)$] do not improve these estimates, furnishing values similar to the ones above.

For $k=4$, the 3-pt extrapolation of the set $(s_2,s_4,s_6)$, with the largest even $L$'s available in this case for the isotropic phase, yields $s_{\infty}(4)=0.09003$. This value lays well at the middle of the interval established in Ref. \cite{gp79} for this entropy in the square lattice. For larger $k$'s, however, we obtain extrapolated results out of the lower and upper bounds from Ref. \cite{gp79}. This is somewhat consistent with the findings from the previous section, pointing that to obtain reliable values of $s_{\infty}(k)$ we should extrapolate data for increasing $L$ as $k$ increases.

\section{Conclusion}
\label{secConc}

We have investigated fully-packed rods on two sequences of generalized Husimi lattices (HLs), which are expected to furnish better approximations for these systems on the square lattice as their level $L$ increases. For $L=1$ we recover the ordinary HL built with elementary squares, whose results are very similar to those previously found for these $k$-mers on the Bethe lattice \cite{drs11}, but with a larger entropy and a smaller nematic order parameter for the nematic phase. On these two lattices, dimers and trimers are found in an isotropic phase, with entropy $s^{(I)}>0$, while for larger rods the stable phase is nematic. On the other hand, our results for the generalized HLs strongly indicate that when $L \rightarrow \infty$ only the isotropic phase is present in the system. In fact, with few exceptions, it is the single phase observed for the higher $L$'s analyzed here. Moreover, the nematic order parameter is a decreasing function of $L$, for a given $k$, when the nematic phase appears. Since we expect to obtain the behavior of the model on the square lattice when $L \rightarrow \infty$, these results are confirming that $k$-mers at full-packing are indeed in an isotropic phase on the square lattice.

The striking agreement of the extrapolated values (for $L \rightarrow \infty$) of the entropy for dimers with the exact result for the square lattice (with a difference of $0.03$\%) confirms that our approaches are indeed a good venue to access the thermodynamic behavior of rods on the regular lattice. Given the difficulties with these extrapolations, due to the $L$-parity effect observed in the entropies, our results for trimers [differing by $3$\% from the best known value of $s(3)$] can be regarded as a very good estimate. Moreover, we obtained also reliable estimates for tetramers on both HLs, whose average yields $s(4) = 0.10(1)$. In general, our results indicate that to obtain accurate estimates, we have to extrapolate data for a maximal level, $L_{max}$, that is larger than and increases with $k$ (i.e., $L_{max} \gtrsim k$). Namely, by increasing the rod size $k$, one should work with building blocks whose size also increases. This is indeed expected and is certainly needed also in other athermal systems (for other particle shapes), as well as in thermal systems with long-range interactions reaching a length $k$. It turns out that, at least for rods, it is quite hard to follow the requirement $L_{max} \gtrsim k$ and, as seen in Tabs. \ref{TabEntDiag} and \ref{TabEntKS}, one rather has $L_{max}$ decreasing with $k$, due to the numerical difficulties in generating and dealing with a large number of recursion relations, each one containing a very large number of terms.

On this matter, we remark that the regular square HLs (introduced by KS) have an advantage over the diagonal square HLs (introduced by Monroe), since in the former case the number of rods' configurations at the root line is much smaller. Thereby, beyond the smaller number of recursion relations, they have much less terms in the KS case, allowing us to study higher levels. Conversely, for a given $k$ and $L$ $(\ge 2)$, the diagonal approach furnishes results closer to the asymptotic ones. In fact, by comparing the data in Tabs. \ref{TabEntDiag} and \ref{TabEntKS}, one sees that $s_L(k)$ is always larger in the diagonal case than in the KS one, with a smaller nematic order parameter and the entropy of the isotropic phase becoming positive at lower $L$'s. This is explained by the effective number of sites in each building block, which is approximately two times greater in the diagonal HLs than in the KS case. So, it is difficult to establish which type of HLs is the best one. For example, the larger number of points to extrapolate in the KS approach yielded better results for the rods for $k \le 4$, but in the diagonal case we obtained a reliable estimate for $s(6)$.

Finally, it is worth discussing what might happen in the more general case where vacancies are also present in the lattice. On the Bethe lattice, this system is found in the isotropic phase at low rod densities, $\rho$, and undergoes a continuous transition to the nematic phase as $\rho$ increases, for $k \ge 4$ \cite{drs11}. This suggests a similar scenario for the cases where the nematic phase was found here and the analysis of the behavior of the (possible) critical points with $L$ is an interesting issue, which might help to explain the disappearance and re-appearance of the nematic phase as $L$ increases, for a given $k$. In fact, for the cases where only the isotropic phase was found here, we may have either the absence of the isotropic-nematic transition (as indeed expected for $k < 7$) or isotropic-nematic-isotropic transitions (as expected for $k \ge 7$ for the square lattice). We are currently initiating the study of this systems, which is much more challenging than the case analyzed here, once the presence of vacancies considerably increases the number of possible configurations and recursion relations to be handled.

\section{Acknowledgements}
We thank R. Rajesh for helpful comments and suggestions and a critical reading of the manuscript. NTR thanks the Brazilian agency CAPES, through the INCT-SC, for financial support. Part of this research utilized Queen Mary's Apocrita HPC facility, supported by QMUL Research-IT. TJO thanks CNPq and FAPEMIG for support.

\appendix

\section{The bulk free energy}
\label{SecAppendix}

Although the bulk free energy \textit{per site}, $\phi_b$, has been derived in
several works, for different HLs (see, e.g., 
\cite{Gujrati,tiagoPol,tiagoALG,Nathann19,Nathann21}),
they are always defined in terms of the \textit{total} free energies 
$\Phi_M = -k_B T \ln Y_{M}$ and $\Phi_{M+1} = -k_B T \ln Y_{M+1}$ for consecutive generations of the tree. In case of a limit cycle of period 2, however, densities  in the system have a layered structure (repeating after each two generations), so that 
it is more appropriate to define $\phi_b$ in terms of $\Phi_{M}$ and $\Phi_{M+n}$, with $n=2,4,6,\ldots$. In 
order to do this, let us start recalling that, by connecting the central plaquettes of adjacent building blocks of the HLs considered here, a Cayley tree with coordination 
$q=4$ is formed [see Fig. \ref{Fight}]. If $V_{eff}$ is the effective number of sites 
in each building block, following Gujrati \cite{Gujrati} the total free energy can be 
written as $\Phi_M = V_{eff} [N_b^{(M)} \phi_b + N_s^{(M)} \phi_s]$, where 
$N_b^{(M)}=2\times 3^{M-1}-1$ and $N_s^{(M)} = 4\times 3^{M-1}$ are the number of 
building blocks in the bulk and at the surface of the HL, respectively, while $\phi_b$ 
and $\phi_s$ are the respective free energy densities there. Then, the bulk free energy per site is given by
\begin{equation}
 \phi_b = \frac{1}{(3^n-1) V_{eff}}\left[\Phi_{M+n} - 3^n \Phi_M \right] = -\frac{k_B 
 T}{(3^n-1) V_{eff}} \ln\left[ \frac{Y_{M+n}}{Y_M^{3^n}} \right].
 \label{eqApp1}
\end{equation}
If one uses $A$ and $B$ to denote each point of the cycle and assumes that generation 
$M$ falls in point $A$, then, for the HL built with diagonal square clusters, $Y_M = 
g_M(0)^4 y_{A}$, $Y_{M+1} = g_{M+1}(0)^4 y_{B}$, $Y_{M+2} = g_{M+2}(0)^4 y_{A}$ and so on. Moreover, for general $n$, one may write
$g_{M+n}(0) = g_M(0)^{3^n} f_A(0)^{P_n} f_B(0)^{P_{n-1}}$, where $P_n = [3^{n+1}+(-1)^{n+1}-2]/8$. Substituting these quantities in Eq. \ref{eqApp1}, considering that $n$ is even, one readily gets
\begin{equation}
 \phi_b = -\frac{k_B T}{V_{eff}}\ln \left[ \frac{f_A(0)^{\frac{3}{2}} f_B(0)^{\frac{1}{2}}}{y_{A}} \right].
 \label{eqApp2}
\end{equation}
For the KS lattice, where one has to distinguish between the ppf's for the $x$ and $y$ directions, a similar derivation yields 
\begin{equation}
 \phi_b = -\frac{k_B T}{V_{eff}}\ln \left[ \frac{f_A(0)^{\frac{3}{4}} f_A^*(0)^{\frac{3}{4}} f_B(0)^{\frac{1}{4}} 
 f_B^*(0)^{\frac{1}{4}}}{y_{A}} \right].
\end{equation}
Note that these free energies are independent of $n$ (even), as expected.
By exchanging $A$ and $B$ in these expressions, one obtains an equivalent definition 
for $\phi_{b}$ [see Eq. \ref{eqEntLqqc2}], which corresponds to the case where 
generation $M$ falls in point $B$.

It is noteworthy that the free energies obtained for odd $n$'s have an unexpected $n$-dependence. For instance, in this case Eq. \ref{eqApp2} changes to
\begin{equation}
 \phi_b = -\frac{k_B T}{V_{eff}} \left\lbrace \ln \left[ \frac{f_A(0)^{\frac{3}{2}} f_B(0)^{\frac{1}{2}}}{y_{A}} \right] + \frac{1}{3^n-1} \ln \left[ \frac{f_A(0) y_B}{f_B(0) y_{A}} \right] \right\rbrace.
 \end{equation}
This confirms that in systems with cycles of period 2 we can not derive a consistent expression for $\phi_b$ from the total free energies for subsequent generations of the tree differing by an odd number.

\end{document}